\documentclass[conference,a4paper]{IEEEtran}

\usepackage{caption}
\IEEEoverridecommandlockouts

\usepackage{algorithm}
\usepackage{algpseudocode}
\usepackage{amsmath}
\usepackage{amsfonts}
\usepackage{graphicx}
\usepackage{epsfig}
\usepackage{cite}
\usepackage[font=small,skip=6pt]{caption}
\usepackage{txfonts}
\usepackage{relsize}
\usepackage{color}
\usepackage{url}
\usepackage{multicol}
\usepackage{float}
\usepackage{lipsum}
\usepackage{multirow}
\usepackage{stfloats}
\usepackage{algorithm}
\usepackage{algorithmicx}
\usepackage{algorithm}
\usepackage{algpseudocode}
\usepackage{amsmath}
\usepackage{graphics}
\usepackage{epsfig}
\usepackage{multicol}
\usepackage{cite}
\usepackage{txfonts}
\usepackage{relsize}
\usepackage{float}
\usepackage{lipsum}
\usepackage{color}
\usepackage[a4paper, top=1.935cm, bottom=4.56 cm, left=1.425cm, right=1.275cm]{geometry}

\ifCLASSINFOpdf
\else
\fi

\begin{document}
\columnsep 0.215in

\title{Downlink Power Allocation for STAR-RIS-Assisted Cell-Free Massive MIMO with Multi-antenna Users
\thanks{This work was supported by the Research Grants Council under the Area of Excellence scheme grant AoE/E-601/22-R.}}
\author{Jun Qian, Ross Murch, and Khaled B. Letaief\\
Dept. of ECE, The Hong Kong University of Science and Technology, Kowloon, Hong Kong
\\Email: eejunqian@ust.hk, eermurch@ust.hk, eekhaled@ust.hk

}

\maketitle
{\begin{abstract}

This paper investigates the downlink power allocation of the simultaneous transmitting and reflecting reconfigurable intelligent surface (STAR-RIS)-assisted cell-free massive multiple-input multiple-output (MIMO) system with multi-antenna users. We introduce downlink spectral efficiency (SE) and derive novel closed-form SE expressions using linear minimum mean squared error (MMSE) detectors. We also address the downlink power allocation via a sum SE maximization problem framed within an alternating direction method of multipliers (ADMM)-based fractional programming (FP) algorithm. Numerical results demonstrate that systems utilizing multi-antenna users significantly enhance SE, achieving at least a $20\%$ SE increase as the number of antennas increases from one to six. Additionally, our proposed ADMM-based FP algorithm outperforms existing fractional power control approaches, yielding a more than $20\%$ SE increase. These results highlight the necessity for adopting multi-antenna users and efficient power allocation algorithms in STAR-RIS-assisted cell-free massive MIMO systems.
\end{abstract} 
 
}

\maketitle

\section{Introduction} 
The advancement of wireless communications has led to significant interest in cell-free massive multiple-input multiple-output (MIMO). This innovative approach effectively reduces inter-cell interference and enhances transmission coverage by combining the strengths of massive MIMO and distributed network architectures \cite{7827017,10297571}. Moreover, reconfigurable intelligent surfaces (RISs) have been proposed to boost spectral efficiency (SE) by introducing cost-effective and energy-efficient controllable cascaded links, especially under harsh propagation conditions\cite{9875036,10167480}.

Recent research has highlighted great interest in wireless communication through RISs. Thus, the integration of RIS and cell-free massive MIMO systems to harness their joint advantages is also pivotal\cite{9875036,10167480}. However, most existing RIS-assisted wireless networks position both receivers and transmitters on the same side of RISs\cite{10297571,10167480,9875036}. In real-world applications, users often appear on both sides of RISs\cite{9570143,10297571}. To address this limitation, \cite{9570143} introduced simultaneous transmitting and reflecting RIS (STAR-RIS) to enable the placement of both receivers and transmitters on either side of the RISs\cite{9570143}. Thus, integrating STAR-RISs and cell-free massive MIMO becomes crucial for enhancing overall system performance\cite{10297571,10841966}. In \cite{10297571}, the authors studied spatially correlated STAR-RIS-assisted cell-free massive MIMO systems. \cite{44444} explored the performance of STAR-RIS-assisted cell-free massive MIMO systems experiencing phase errors and electromagnetic interference, while \cite{10841966} focused on scenarios with hardware impairments. \cite{10264149} investigated the SE performance of active STAR-RIS-assisted cell-free massive MIMO systems. Meanwhile, \cite{44444,10297571} emphasized that introducing STAR-RISs in cell-free massive MIMO systems outperforms that with conventional RISs.

To the best of our knowledge, existing literature predominantly focuses on single-antenna users within STAR-RIS-assisted cell-free massive MIMO \cite{10297571,10841966,10264149,44444}. However, contemporary users are expected to employ multiple antennas enabled by advanced antenna technology \cite{7500452}. Introducing multi-antenna users in cell-free massive MIMO systems can lead to substantial SE increases with refined power control \cite{9079911} and compensate for the adverse effects of hardware impairments \cite{10163977,10201892}. Meanwhile, although power allocation is well-explored in cell-free massive MIMO systems \cite{9293031}, STAR-RIS-assisted cell-free massive MIMO presents a different landscape due to a larger number of optimization variables and a different problem structure. Thus, \cite{10841966} studied the max-min fairness power control and \cite{44444} investigated the fractional power control in STAR-RIS-assisted cell-free massive MIMO.
Despite the above potentials, limited work has focused on the analysis of STAR-RIS-assisted networks with multi-antenna users \cite{qian2025starrisassisted}, and efficient power allocation is necessary for performance improvement. To bridge these gaps in the literature, we are motivated to analyze the performance 
of STAR-RIS-assisted cell-free massive MIMO with multi-antenna users facilitated with efficient downlink power allocation.

This work is the first to analyze STAR-RIS-assisted cell-free massive MIMO systems with multi-antenna users. It introduces an efficient low-complexity power allocation based on the alternating direction method of multipliers (ADMM) and fractional programming (FP) algorithm \cite{9293031,7874154}. Thus, Section II presents the STAR-RIS-assisted cell-free massive MIMO system model with multi-antenna users on the linear minimum mean squared error (MMSE) detectors, deriving novel closed-form expressions. Section III details the ADMM-based FP algorithm by leveraging downlink sum SE maximization and analyzes the computational complexity. Section IV presents numerical results demonstrating that increasing the number of user antennas can enhance the SE performance, and the proposed downlink power allocation exhibits significant performance improvements. Finally, Section V concludes the work.
\begin{figure}[t!]
		\centering
	\includegraphics[width=0.76\columnwidth]{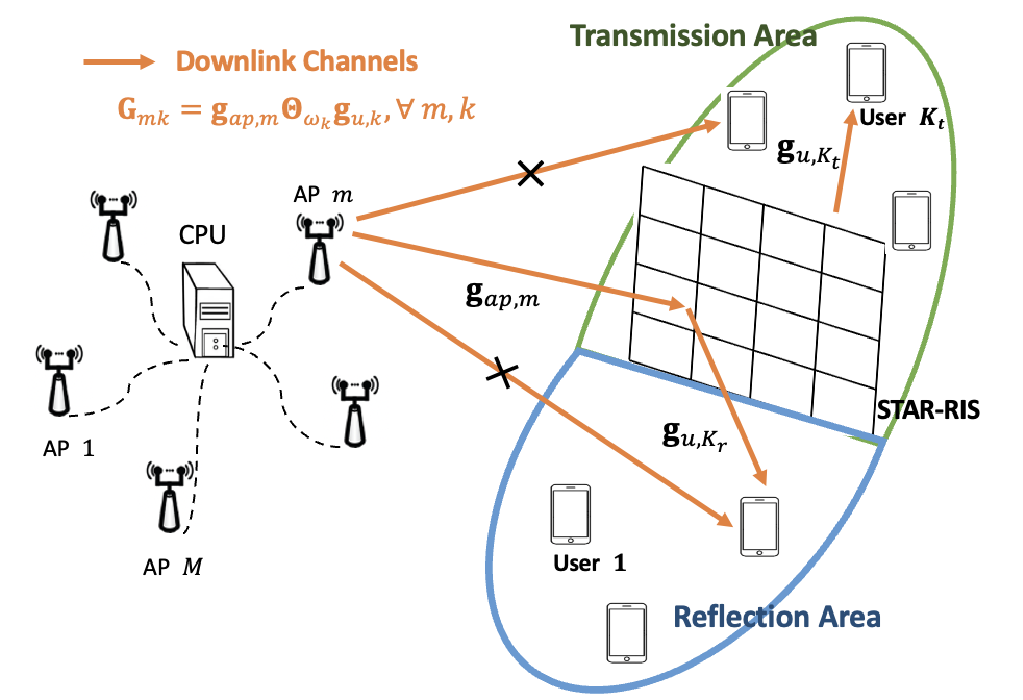} 
        \caption{Illustration of STAR-RIS-assisted cell-free massive MIMO systems.}
		\label{System}
          \vspace{-5pt}
        \end{figure}

\section{System Model}
We consider a STAR-RIS-assisted cell-free massive MIMO system, where $M$ $N_{ap}$-antenna APs connected to the CPU via ideal fronthaul links serve $K$ $N_{u}$-antenna users simultaneously\cite{10297571,10264149}.
The $L$-element STAR-RIS facilitates communication between APs and users. The reflection-area users are positioned on the same side as the APs and indexed by the set $\mathcal{K}_r$, with a set cardinality of $|\mathcal{K}_r|=K_r$. The transmission-area users are located on the opposite side as APs and indexed by set $\mathcal{K}_t$ with $|\mathcal{K}_t|=K_t$ ( $K_r+K_t=K$ and $\mathcal{K}_r\cap\mathcal{K}_t=\varnothing$). We use $\omega_k,~\forall k$ to denote the STAR-RIS operation mode for the $k$-th user and $\mathcal{W}_k$ as the set of users sharing the same operation mode as the $k$-th user, with $\omega_k=r$ for the $k$-th reflection-area user and $\omega_k=t$ for the $k$-th transmission-area user \cite{10297571,44444}. 
\addtolength{\rightmargin}{0.02in}

\subsection{STAR-RIS-Assisted Channel Model}
The proposed system operates in time-division duplex (TDD) mode under a harsh propagation environment,
where the direct links between APs and users are obstructed due to multi-scatterer distribution \cite{10297571}. The associated Rayleigh-fading channels are described by the classical Kronecker model\cite{10163977}. The Rician-distributed channels are left for future work. Therefore, as shown in Fig. \ref{System}, the STAR-RIS-assisted uplink channel between the $k$-th user to the $m$-th AP, $\textbf{G}_{mk}\in \mathbb{C}^{N_{ap}\times N_{u}}$, can be formulated as
\begin{equation}
     \displaystyle \textbf{G}_{mk}=
\underbrace{\sqrt{\beta_{ap,m}}\textbf{v}_{ap,m}{\textbf{R}}_{m,t}^{{1/2}}}_{\textbf{g}_{ap,m}}\boldsymbol{\Theta}_{\omega_k}\underbrace{\sqrt{\beta_{u,k}}{\textbf{R}}_{k,r}^{1/2}\textbf{v}_{u,k}}_{\textbf{g}_{u,k}},
     \label{aggregate_uplink_channel}
   \end{equation}
where $\textbf{g}_{ap,m}\in\mathbb{C}^{N_{ap}\times L}$ is the channel from the STAR-RIS to the $m$-th AP and $\textbf{g}_{u,k}\in\mathbb{C}^{L\times N_{u}}$ is the channel from the $k$-th user to the STAR-RIS.
$\beta_{ap,m}$ denotes the large-scale fading coefficient between the $m$-th AP and the STAR-RIS, and $\beta_{u,k}$ is that between the $k$-th user and the STAR-RIS. $\textbf{v}_{ap,m}\in \mathbb{C}^{N_{ap}\times L}$ and $\textbf{v}_{u,k}\in \mathbb{C}^{L\times N_u}$ are independent fast-fading channels with $\mathcal{CN}(0,1)$ independent and identically distributed random variables.
   ${\textbf{R}}_{m,t}={\textbf{R}}_{k,r} =A{\textbf{R}} \in \mathbb{C}^{L\times L}$ denotes the STAR-RIS spatial correlation matrix. The STAR-RIS element area is $A=d_Hd_V$, where $d_H$ and $d_V$ are the respective horizontal width and vertical height\cite{9598875}. The $(x,y)$-th element in $\textbf{R}$ is $[\textbf{R}]_{x,y}=\text{sinc}\Big{(}\frac{2||\textbf{u}_x-\textbf{u}_y||}{\lambda}\Big{)}$,
with sinc function $\text{sinc}(a)=\text{sin}(\pi a)/(\pi a)$ and carrier wavelength $\lambda$. $\textbf{u}_x=[0,\text{mod}(x-1,L_h)d_h,\lfloor(x-1)/L_h\rfloor d_v]^T$ denotes the position vector\cite{9598875,44444}, where $L=L_h\times L_v$ with $L_h$ and $L_v$ being the numbers of column and row elements at STAR-RIS. We assume the STAR-RIS follows the energy splitting (ES) protocol, where the STAR-RIS coefficient matrices are $\boldsymbol{\Theta}^{}_{r}=\text{diag}(u_{1}^r{\theta}_{1}^r,u_{2}^r{\theta}_{2}^r,...,u_{L}^r{\theta}_{L}^r)\in\mathbb{C}^{L\times L}$ for the reflection mode and $\boldsymbol{\Theta}^{}_{t}=\text{diag}(u_{1}^t{\theta}_{1}^t,u_{2}^t{\theta}_{2}^t,...,u_{L}^t{\theta}_{L}^t)\in\mathbb{C}^{L\times L}$ for the transmission mode, with the induced phase shifts ${\theta}_{l}^t=e^{i\varphi_{l}^t},~{\theta}_{l}^r=e^{i\varphi_{l}^r}$, $\varphi_{l}^t,~\varphi_{l}^r \in[0,2\pi)$ and the amplitude coefficients $u_{l}^t,~u_{l}^r \in[0,1]$, $(u_{l}^t)^2+(u_{l}^r)^2=1$, $\forall l$ \cite{10297571,10264149}. 
Then, the vectorization form of $\textbf{G}_{mk}$, $\textbf{g}_{mk}=\text{vec}(\textbf{G}_{mk})$, is given by
\begin{flalign}
& 
\displaystyle \textbf{g}_{mk}\displaystyle=\sqrt{\beta_{ap,m}\beta_{u,k}}\left(\textbf{I}_{N_u}\otimes\textbf{v}_{ap,m}\right)\left(\textbf{I}_{N_u}\otimes(A{\textbf{R}}^{1/2}\boldsymbol{\Theta}_{\omega_k}{\textbf{R}}^{1/2})\right)\text{vec}(\textbf{v}_{u,k}).
     \label{channel_user_RIS_vec}
   \end{flalign}
 Accordingly, the aggregate uplink channel follows $\textbf{G}_{mk}\sim\mathcal{CN}(\textbf{0},\mathbf{\Delta}_{mk})$, where the covariance matrix is $\mathbf{\Delta}_{mk}=\mathbb{E}\{\textbf{g}_{mk}(\textbf{g}_{mk})^H\}\displaystyle=\bar{{\Delta}}_{mk}\left({\textbf{I}}_{N_u} \otimes{\textbf{I}}_{N_{ap}}\right)\in\mathbb{C}^{N_{ap}N_u\times N_{ap}N_u}$, with $\bar{{\Delta}}_{mk}=\beta_{ap,m}\beta_{u,k}\text{tr}(\textbf{T}_{\omega_k})$ and ${\textbf{T}}_{\omega_k}\displaystyle=A^2{\textbf{R}}^{1/2}\boldsymbol{\Theta}_{\omega_k}\textbf{R}\boldsymbol{\Theta}_{\omega_k}^H{\textbf{R}}^{1/2}$.

\subsection{Uplink Channel Estimation}
For uplink channel estimation, $\tau_p$ mutually orthogonal pilot matrices are transmitted by users to all APs\cite{10201892}, and each pilot matrix contains $N_u$ $\tau_p$-length pilot sequences selected from the pilot book.
We define $\mathbf{\Phi}_{k}\in\mathbb{C}^{\tau_p\times N_u}$ as the pilot matrix for $k$-th user, where $\mathbf{\Phi}_{k}^H\mathbf{\Phi}_{k'}=\textbf{I}_{N_u},~k'\in\mathcal{P}_k$, otherwise equals to $\mathbf{0}_{N_u}$. 
Since pilot resources are limited in practice, we define $\mathcal{P}_k$ as the index subset of users that uses the same pilot matrix as the $k$-th user, including itself \cite{9737367,10201892}.
Thus, the received signal of the $m$-th AP, $\textbf{Y}_{m,p}\in\mathbb{C}^{N_{ap}\times \tau_p}$, is expressed as
\begin{equation}
\begin{array}{ll}
     \displaystyle \textbf{Y}_{m,p}      \displaystyle =\sum\nolimits_{k=1}^{K}\textbf{G}_{mk}\sqrt{\tau_pp_{p}\xi_k}\mathbf{\Phi}_{k}^{H}\displaystyle+\textbf{N}_{m,p},
\end{array}
\label{Uplink_Received_Pilot}
   \end{equation}
where $p_p$ denotes the maximum pilot power of each user, $\xi_k$ represents the pilot power control coefficient for each antenna of the $k$-th user with $N_{u}\xi_{k}\leq 1$ \cite{10163977,10201892,9737367}. For simplicity, we apply $\xi_{k}=1/N_u,\forall k$. Moreover, $\textbf{N}_{m,p}\in\mathbb{C}^{N_{ap}\times \tau_p}$ stands for the additive white Gaussian noise (AWGN) matrix at the $m$-th AP, where the $v$-th column satisfies $\Big{[}\textbf{N}_{m,p}\Big{]}_v\sim \mathcal{CN}(\textbf{0},\sigma^2\textbf{I}_{N_{ap}})$. 

To obtain accurate channel estimation, the $m$-th AP multiplies $\textbf{Y}_{m}$ with $\mathbf{\Phi}_{k}$ to despread the received signal. We introduce the vectorization form, $\textbf{y}_{mk,p}=\text{vec}(\textbf{Y}_{m}\mathbf{\Phi}_{k})$, as 
 \begin{equation}
\begin{array}{ll}
     \displaystyle
\textbf{y}_{mk,p}\displaystyle=\sum\nolimits_{k'\in\mathcal{P}_k}\sqrt{\tau_pp_{p}\xi_{k'}}\textbf{g}_{mk'}\vspace{1 pt}\displaystyle+\left(\mathbf{\Phi}_{k}^T\otimes\textbf{I}_{N_{ap}}\right)\text{vec}(\textbf{N}_{m,p}).
\end{array}
\label{Pilot_projection_vectorized}
   \end{equation}
Subsequently, the linear MMSE channel estimation of ${\textbf{G}}_{mk}$ in the form of vectorization \cite{10163977,9079911} is given by 
  \begin{equation}
\begin{array}{ll}
     \displaystyle \hat{\textbf{g}}_{mk}=\text{vec}(\hat{\textbf{G}}_{mk})=\bar{z}_{mk}\textbf{y}_{mk,p}.
\end{array}
   \end{equation}
Consequently, the estimated channel $\hat{\textbf{G}}_{mk}
$
is distributed as $\mathcal{CN}\left(\textbf{0},\hat{\mathbf{\Delta}}_{mk}\right)$, 
where $   \displaystyle \hat{\mathbf{\Delta}}_{mk}\displaystyle=z_{mk}\left(
\textbf{I}_{N_u}\otimes\textbf{I}_{N_{ap}}\right),$
with $ \displaystyle z_{mk}=\sqrt{\tau_pp_p\xi_k}\bar{\Delta}_{mk}\bar{z}_{mk}$ and $ \displaystyle\bar{z}_{mk}=\frac{\sqrt{\tau_pp_p\xi_k}\bar{\Delta}_{mk}}{\sum\nolimits_{k'\in\mathcal{P}_k}\tau_pp_{p}\xi_{k'}\bar{{\Delta}}_{mk'}+\sigma^2}$.

\subsection{Downlink Data Transmission and SE Analysis}
For downlink data transmission, we assume that each AP adopts the conjugate beamforming ${\textbf{w}}_{mk}=\hat{\textbf{G}}_{mk}$ to precode the desired signal\cite{9079911,44444}. Subsequently, the preceded signal will be transmitted to all users. The transmitted signal $\textbf{x}_m\in\mathbb{C}^{N_{ap}\times 1}$ at the $m$-th AP is $\textbf{x}_m=\sum\nolimits_{k=1}^{K}\sqrt{\eta_{mk}}\textbf{w}_{mk}\textbf{q}_{k}$,
where $\textbf{q}_{k}$ with $\mathbb{E}\{\textbf{q}_{k}\textbf{q}_{k}^H\}=\textbf{I}_{N_u}$ is the symbol vector intended for the $k$-th user. $\eta_{mk}$ is the power control coefficient satisfying $\mathbb{E}\{||\textbf{x}_m||^2\}=\sum\nolimits_{k=1}^K\eta_{mk}\text{tr}\left(\hat{\mathbf{\Delta}}_{mk}\right)
\leq p_d,~\forall m$, where $p_d$ is the downlink transmit power \cite{9737367}.
The received signal, $\textbf{r}_k\in\mathbb{C}^{N_u\times 1}$ at the $k$-th user can be given by
 \begin{equation}
\begin{array}{ll}
     \displaystyle \textbf{r}_k&\displaystyle=\sum\nolimits_{m=1}^M\textbf{G}_{mk}^H\textbf{x}_m+\textbf{n}_k\vspace{1 pt}\\&\displaystyle=
     \sum\nolimits_{m=1}^M\sqrt{\eta_{mk}}\mathbb{E}\{\textbf{G}_{mk}^H\textbf{w}_{mk}\}\textbf{q}_{k}\\&\displaystyle+\sum\nolimits_{m=1}^M\sqrt{\eta_{mk}}\left(\textbf{G}_{mk}^H\textbf{w}_{mk}-\mathbb{E}\{\textbf{G}_{mk}^H\textbf{w}_{mk}\}\right)\textbf{q}_{k}\vspace{1 pt}\\&\displaystyle+
\sum\nolimits_{k'\neq k}\sum\nolimits_{m=1}^M\sqrt{\eta_{mk'}}\textbf{G}_{mk}^H\textbf{w}_{mk'}\textbf{q}_{k'}+\textbf{n}_k,
     
\end{array}\label{r_k}
   \end{equation}
where $\textbf{n}_k\in\mathbb{C}^{N_u\times 1}$ is the noise vector following $\mathcal{CN}(0,\sigma^2\textbf{I}_{N_u})$.

\textit{Theorem 1:} We deliver the achievable downlink SE of the $k$-th user by linear MMSE detection with the linear MMSE detector $\textbf{f}_{k,n}\in\mathbb{C}^{N_u\times 1}$ for the $n$-th data stream of the $k$-th user
\begin{equation}
\begin{array}{ll}
     \displaystyle \textbf{f}_{k,n}=\left(\sum\nolimits_{k'=1}^K\mathbb{E}\Big{\{} \textbf{C}_{kk'}\textbf{C}_{kk'}^H\Big{\}}
+\mathbb{E}\Big{\{}\textbf{n}_k\textbf{n}_k^H\Big{\}}\right)\textbf{d}_{kn},
\end{array}\label{Linear_detector}
   \end{equation}
  where $\textbf{d}_{kn}$ is the $n$-th column of $\textbf{D}_{k}\in\mathbb{C}^{N_u\times N_u}$ \cite{7500452,9079911}. $\textbf{D}_{k}=\sum\nolimits_{m=1}^M\sqrt{\eta_{mk}}\mathbb{E}\Big{\{}\textbf{G}_{mk}^H\textbf{w}_{mk}\Big{\}}\in\mathbb{C}^{N_u\times N_u}$,  
$\textbf{C}_{kk'}=\sum\nolimits_{m=1}^M\sqrt{\eta_{mk'}}\textbf{G}_{mk}^H\textbf{w}_{mk'}$. 
Given MMSE detectors, \eqref{Linear_SE} expresses the downlink SE of the $k$-th user with the closed-form expressions at the top of this page with assistance of $\bar{D}_{kn}=\sum\nolimits_{m=1}^M\sqrt{\eta_{mk}}z_{mk}N_{ap}$
and \eqref{Sigma_k_closed_form} formulates $\bar{C}_{kn}$ at the top of this page.
     \begin{figure*}
\begin{equation}
\begin{array}{ll}
     \displaystyle \text{SE}_{k}^\text{LINEAR}\displaystyle=\sum\nolimits_{n=1}^{N_u}\mathbb{E}\left\{\text{log}_2\Bigg{(}1+\frac{\big{|}\textbf{f}_{k,n}^H\textbf{d}_{kn}\big{|}^2}{\textbf{f}_{k,n}^H \left( \displaystyle\sum\nolimits_{k'=1}^K\mathbb{E}\Big{\{} \textbf{C}_{kk'}\textbf{C}_{kk'}^H\Big{\}}
+\mathbb{E}\Big{\{}\textbf{n}_k\textbf{n}_k^H\Big{\}}\right)\textbf{f}_{k,n}-\big{|}\textbf{f}_{k,n}^H\textbf{d}_{kn}\big{|}^2}\Bigg{)}\right\}=\displaystyle
\sum\nolimits_{n=1}^{N_u}\text{log}_2\left(1+\frac{\bar{D}_{kn}^2}{\bar{C}_{kn}-\bar{D}_{kn}^2+\sigma^2}\right)
.
\end{array}\label{Linear_SE}
 \vspace{-4 pt}
   \end{equation}
       \vspace{-10 pt}
\hrulefill
     \end{figure*}  
  \begin{figure*}[t!]
\begin{equation}
\begin{array}{ll}
\displaystyle\bar{C}_{kn}&\displaystyle=
\sum\nolimits_{k'\in\mathcal{P}_k}\sum\nolimits_{m=1}^M\eta_{mk'}\bar{z}_{mk'}^2\tau_pp_p\xi_{k}
\left(
     \beta_{ap,m}^2\beta_{u,k}^2\text{tr}(\textbf{T}_{\omega_{k}}^2)N_{ap}
     \right)\displaystyle
 \displaystyle\displaystyle+\sum\nolimits_{k'=1}^K\sum\nolimits_{m=1}^M\eta_{mk'}\bar{z}_{mk'}^2N_uN_{ap}\bar{\Delta}_{mk}\left(
\sum\nolimits_{k''\in\mathcal{P}_{k'}}\tau_pp_p\xi_{k''}\bar{\Delta}_{mk''}
     +
     \sigma^2
     \right)\\&\displaystyle
     +   \sum\nolimits_{k'\in\mathcal{P}_k}\sum\nolimits_{m=1}^M\sum\nolimits_{m'=1}^M\sqrt{\eta_{mk'}\eta_{m'k'}}\bar{z}_{mk'}\bar{z}_{m'k'}\tau_pp_p\xi_k\bar{\Delta}_{mk}\bar{\Delta}_{m'k}N_{ap}^2\\&\displaystyle\displaystyle
     +\sum\nolimits_{k'=1}^K\sum\nolimits_{m=1}^M\sum\nolimits_{m'=1}^M
     \sqrt{\eta_{mk'}\eta_{m'k'}}\bar{z}_{mk'}\bar{z}_{m'k'}N_u\left(\sum\nolimits_{k''\in\mathcal{P}_{k'}}\tau_pp_p\xi_{k''}\beta_{ap,m}\beta_{ap,m'}\beta_{ap,k}\beta_{ap,k''}\text{tr}(\textbf{T}_{\omega_{k}}\textbf{T}_{\omega_{k''}})
     N_{ap}^2
     \right).
\end{array}\label{Sigma_k_closed_form}
   \vspace{-4 pt}
   \end{equation}
       \vspace{ -8 pt}
   \hrulefill
   \end{figure*}
   
\textit{Theorem 2:} The downlink SE of the $k$-th user with MMSE-based successive interference cancellation (SIC) schemes \cite{7500452,9079911} and its closed-form expressions are given by
\begin{equation}
\begin{array}{ll}
     \displaystyle \text{SE}^\text{MMSE-SIC}_{k}&\displaystyle=
     \text{log}_2\text{det}\left(\textbf{I}_{N_u}+\textbf{D}_{k}^H\mathbf{\Sigma}_{k}^{-1}\textbf{D}_{k}\right)\\&\displaystyle=
     \text{log}_2\text{det}\left(\textbf{I}_{N_u}+\frac{\bar{D}_k^2}{\bar{C}_k-\bar{D}_k^2+\sigma^2}\textbf{I}_{N_u}\right).
\end{array}\label{SE_MMSE_SIC}
   \end{equation}
It is evident that system performance is identical regardless of MMSE-SIC or linear MMSE detectors under given scenarios\cite{9079911}. For simplicity, this work focuses on the linear MMSE detector. The spatial correlation and mutual coupling of multi-antenna AP and user structures are left for future work.
   
\section{Downlink Sum SE Maximization}
In this section, we study the downlink sum SE maximization by optimizing the power control coefficients, leveraging the ADMM-based FP algorithm in \cite{9293031}. The proposed system differs from \cite{9293031} due to the application
of STAR-RIS and multi-antenna users, introducing a different problem structure with multiple optimization variables. 
\subsection{ADMM-based FP Algorithm}
For the proposed power control optimization, we introduce an alternative variable $\zeta_{mk}=\sqrt{\eta_{mk}z_{mk}}$ with $\boldsymbol{\zeta}_k=[\zeta_{1k},...,\zeta_{Mk}]^T\in\mathbb{C}^{M\times 1}$, $\boldsymbol{\zeta}=[\zeta_{11},...,\zeta_{MK}]^T\in\mathbb{C}^{MK\times 1}$ to express the
SINR as a ratio: $\text{SINR}_{kn}={A_{kn}(\boldsymbol{\zeta})}/{B_{kn}(\boldsymbol{\zeta})}$. Then, the power control optimization problem can be given by \cite{10517887,9519163,9293031}
  \begin{subequations}
  \begin{align}
   P_1:~&\mathop {\max }\limits_{\boldsymbol{\zeta}}~\sum\nolimits_{k=1}^K\sum\nolimits_{n=1}^{N_u}\text{log}_2\left(1+\frac{A_{kn}(\boldsymbol{\zeta})}{B_{kn}(\boldsymbol{\zeta})}\right)
   \\
   & \text{subject~to} \nonumber \\
   & \sum\nolimits_{k=1}^K \zeta_{mk}^2 \leq \frac{p_d}{N_{ap}N_u},~\forall m.
  \end{align}
  \label{GA_optimization2}
 \end{subequations}
Here, we introduce $A_{kn}(\boldsymbol{\zeta})=\boldsymbol{\zeta}_k^T\textbf{a}_{kn}\textbf{a}_{kn}^T\boldsymbol{\zeta}_k$ and $B_k(\boldsymbol{\zeta})$ are non-negative real function and
positive real function of $\boldsymbol{\zeta}$ with
   \begin{equation}
\begin{array}{ll}
   \displaystyle B_{kn}(\boldsymbol{\zeta})&\displaystyle=\sum\nolimits_{k'\in\mathcal{P}_k}\boldsymbol{\zeta}_{k'}^T\left(\textbf{K}_{kk'n}^{(1)}+\textbf{K}_{kk'n}^{(3)}\right)\boldsymbol{\zeta}_{k'}\\&\displaystyle+\sum\nolimits_{k'=1}^K\boldsymbol{\zeta}_{k'}^T\left(\textbf{K}_{kk'n}^{(2)}+\textbf{K}_{kk'n}^{(4)}\right)\boldsymbol{\zeta}_{k'}-A_{kn}(\boldsymbol{\zeta})+\sigma^2,
\end{array}\label{B_k}
  \end{equation}
   where $\textbf{a}_{kn}\in\mathbb{C}^{M\times 1}$ with $a_{mkn}=\sqrt{z_{mk}}N_{ap}$, $\textbf{K}_{kk'n}^{(1)},~ \textbf{K}_{kk'n}^{(2)}\in\mathbb{C}^{M\times M}$ are diagonal matrices and $\textbf{K}_{kk'n}^{(3)}, ~\textbf{K}_{kk'n}^{(4)}\in\mathbb{C}^{M\times M}$ with 
\begin{flalign}
& 
\displaystyle[\textbf{K}_{kk'n}^{(1)}]_{mm}=\frac{\bar{z}_{mk'}}{\sqrt{\tau_pp_p\xi_{k'}}\bar{\Delta}_{mk'}}\tau_pp_p\xi_{k}
     \beta_{ap,m}^2\beta_{u,k}^2\text{tr}(\textbf{T}_{\omega_{k}}^2)N_{ap},
&\\&
\displaystyle[\textbf{K}_{kk'n}^{(2)}]_{mm}=\frac{\bar{z}_{mk'}N_uN_{ap}\bar{\Delta}_{mk}}{\sqrt{\tau_pp_p\xi_{k'}}\bar{\Delta}_{mk'}}\left(
\sum\nolimits_{k''\in\mathcal{P}_{k'}}\tau_pp_p\xi_{k''}\bar{\Delta}_{mk''}
     +
     \sigma^2
     \right),
   &\\&
\displaystyle[\textbf{K}_{kk'n}^{(3)}]_{mm'}=\frac{\sqrt{{z}_{mk'}{z}_{m'k'}}}{\xi_{k'}\bar{\Delta}_{mk'}\bar{\Delta}_{m'k'}}\xi_k\bar{\Delta}_{mk}\bar{\Delta}_{m'k}N_{ap}^2,
      &\\&
\displaystyle[\textbf{K}_{kk'n}^{(4)}]_{mm'}\displaystyle =\frac{\sqrt{{z}_{mk'}{z}_{m'k'}}N_u N_{ap}^2}{\xi_{k'}\bar{\Delta}_{mk'}\bar{\Delta}_{m'k'}}\times&\\\nonumber&\displaystyle ~~~~~~~~~~~\sum\nolimits_{k''\in\mathcal{P}_{k'}}\xi_{k''}\beta_{ap,m}\beta_{ap,m'}\beta_{u,k}\beta_{u,k''}\text{tr}(\textbf{T}_{\omega_{k}}\textbf{T}_{\omega_{k''}}).
\end{flalign}
   Thus, they satisfy the conditions required for the general FP algorithm \cite{9293031,9519163} and we can express $P_1$ in equivalent Lagrange dual transform as
 \begin{subequations}
  \begin{align}
   P_2:~&\mathop {\max }\limits_{\boldsymbol{\zeta},\boldsymbol{\gamma}}~f_1(\boldsymbol{\zeta},\boldsymbol{\gamma})
   \\
   & \text{subject~to} \nonumber \\
   & \sum\nolimits_{k=1}^K \zeta_{mk}^2 \leq \frac{p_d}{N_{ap}N_u},~\forall m,
  \end{align}
  \label{GA_optimization3}
 \end{subequations}where $\boldsymbol{\gamma}=[\gamma_{11},...,\gamma_{KN_u}]^T\in\mathbb{R}^{KN_u\times 1}$ represents the set of auxiliary SINR variables and the new objective function generated by the Lagrangian dual
transform \cite{10517887,9519163} is
 \begin{equation}
\begin{array}{ll}
   \displaystyle f_1(\boldsymbol{\zeta},\boldsymbol{\gamma})=\sum\nolimits_{k=1}^K\sum\nolimits_{n=1}^{N_u}\left(\text{log}(1+\gamma_{kn})-\gamma_{kn}+\frac{\displaystyle (1+\gamma_{kn})A_{kn}(\boldsymbol{\zeta})}{A_{kn}(\boldsymbol{\zeta})+B_{kn}(\boldsymbol{\zeta})}\right).
\end{array}\label{f_1}
  \end{equation}
When $\boldsymbol{\zeta}$ is held fixed, we can find the local optimum of $\boldsymbol{\gamma}$ by
solving $\partial f_1(\boldsymbol{\zeta},\boldsymbol{\gamma})/{\partial \gamma_{kn}}=0$ to obtain the optimal $\gamma_{kn}^*=\frac{A_{kn}(\boldsymbol{\zeta})}{B_{kn}(\boldsymbol{\zeta})}$ \cite{9293031}. When $\gamma_{kn}, ~\forall k,n$ are fixed, 
only the last term in $f_1(\boldsymbol{\zeta},\boldsymbol{\gamma})$ depends on $\boldsymbol{\zeta}$. We utilize Quadratic Transform \cite{9293031} to reformulate the sum-of-ratio problems, and \eqref{f_2} expresses the new objective function at the top of this page, 
\begin{figure*}[t!]
\begin{equation}
\begin{array}{ll}
   \displaystyle f_2(\boldsymbol{\zeta},\boldsymbol{\varpi})\displaystyle=\sum\limits_{k=1}^K\sum\limits_{n=1}^{N_{u}}\left(2\varpi_{kn}\sqrt{(1+\gamma_{kn})A_{kn}(\boldsymbol{\zeta})}-\varpi_{kn}^2\left(A_{kn}(\boldsymbol{\zeta})+B_{kn}(\boldsymbol{\zeta})\right)\right)\displaystyle=\sum\limits_{k=1}^K\sum\limits_{n=1}^{N_{u}}\left(2\varpi_{kn}\sqrt{(1+\gamma_{kn})}\textbf{a}_{kn}^T\boldsymbol{\zeta}_k-\varpi_{kn}^2\left(A_{kn}(\boldsymbol{\zeta})+B_{kn}(\boldsymbol{\zeta})\right)\right),
\end{array}\label{f_2}
\vspace{-10 pt}
  \end{equation}
  \vspace{-10 pt}
  \hrulefill
  \end{figure*}
in which $\boldsymbol{\varpi}=[\varpi_{11},...,\varpi_{KN_u}]^T\in\mathbb{R}^{KN_u\times 1}$ is a new set of auxiliary variables with $\boldsymbol{\gamma}$ kept constant as $\gamma_{kn}^*=\frac{A_{kn}(\boldsymbol{\zeta})}{B_{kn}(\boldsymbol{\zeta})},~\forall k$. Similarly, we can
obtain the optimal auxiliary variable in $\boldsymbol{\varpi}$ by holding $\boldsymbol{\zeta}$ constant and solving for $\partial f_2(\boldsymbol{\zeta},\boldsymbol{\varpi})/{\partial \varpi_{kn}}=0$ to have the optimum of $\varpi_{kn}^*=\frac{ \sqrt{(1+\gamma_{kn})A_{kn}(\boldsymbol{\zeta})}}{ A_{kn}(\boldsymbol{\zeta})+B_{kn}(\boldsymbol{\zeta})},~\forall k,n$. 
Note that $f_2(\boldsymbol{\zeta},\boldsymbol{\varpi})$ is a concave function, and the resulting optimization problem is treated as a quadratically-constrained quadratic program with $\boldsymbol{\zeta}$. Then, we follow the ADMM-based reformulation to find closed-form update equations for $\{\zeta_{mk}\}$\cite{9293031,7874154}. We reformulate the problem with fixed $\{\gamma_{kn},\varpi_{kn}\}$ by introducing new auxiliary variables $\{q_{mk}\}$
 \begin{subequations}
  \begin{align}
   P_3:~&\mathop {\min}\limits_{\boldsymbol{\zeta},\boldsymbol{q}}~\mathbb{I}_{\mathcal{C}}(\boldsymbol{q})-f_2(\boldsymbol{\zeta},\boldsymbol{\varpi})
   \\
   & \text{subject~to} \nonumber \\
   & \boldsymbol{q}=\boldsymbol{\zeta},
  \end{align}
  \label{GA_optimization4}
 \end{subequations}where $\boldsymbol{q}=[q_{11},...,q_{MK}]^T\in\mathbb{R}^{MK\times 1}$ and $\mathcal{C}$ is the feasible region of the constraint (20b). Meanwhile, the indicator function is defined as $\mathbb{I}_{\mathcal{C}}(\boldsymbol{q})=0$ when $ \sum\nolimits_{k=1}^K q_{mk}^2 \leq \frac{p_d}{N_{ap}N_u},~\forall m$; otherwise, $\mathbb{I}_{\mathcal{C}}(\boldsymbol{q})=+\infty$. Then, we utilize $u_{mk}$ in $\boldsymbol{u}=[u_{11},...,u_{MK}]^T\in\mathbb{R}^{MK\times 1}$ to denote the scaled dual variables for the equality constraint $q_{mk}=\zeta_{mk},~\forall m,k$ in (20b). Following the ADMM methodology \cite{9293031,7874154}, we can transform $P_3$ into
\begin{equation}
\begin{array}{ll}
   \displaystyle f_L(\boldsymbol{\zeta},\boldsymbol{q},\boldsymbol{u})=\mathbb{I}_{\mathcal{C}}(\boldsymbol{q})-f_2(\boldsymbol{\zeta},\boldsymbol{\varpi})\displaystyle+\frac{\xi}{2}\sum\limits_{m=1}^K\sum\limits_{k=1}^K(q_{mk}-\zeta_{mk}+u_{mk})^2,
\end{array}\label{f_L}
  \end{equation}
where $\xi>0$ is the penalty parameter. Then, we can find the update of $\boldsymbol{\zeta}$ as
\begin{equation}
\begin{array}{ll}
   \displaystyle \boldsymbol{\zeta}\displaystyle\gets\text{arg}\min\limits_{\boldsymbol{\zeta}}-f_2(\boldsymbol{\zeta},\boldsymbol{\varpi})+\frac{\xi}{2}\sum\nolimits_{m=1}^M\sum\nolimits_{k=1}^K(q_{mk}-\zeta_{mk}+u_{mk})^2.
   
\end{array}\label{zeta_update}
  \end{equation}
By equating the
gradient of \eqref{zeta_update} to zero, we can acquire the closed-form solution of $\boldsymbol{\zeta}_k$ as 
\begin{equation}
\begin{array}{ll}
   \displaystyle \boldsymbol{\zeta}_k=\bar{\textbf{A}}_k^{-1}\bar{\textbf{c}}_k,~\forall k,
\end{array}\label{zeta_update2}
  \end{equation}
  where 
\begin{equation}
\begin{array}{ll}
   \displaystyle \bar{\textbf{A}}_k&\displaystyle=\sum\nolimits_{k'\in\mathcal{P}_{k}}\sum\nolimits_{n=1}^{N_u}\varpi_{k'n}^2\left(\textbf{K}_{k'kn}^{(1)}+\textbf{K}_{k'kn}^{(3)}\right)\\&\displaystyle+\sum\nolimits_{k'=1}^K\sum\nolimits_{n=1}^{N_u}\varpi_{k'n}^2\left(\textbf{K}_{k'kn}^{(2)}+\textbf{K}_{k'kn}^{(4)}\right)+\frac{\xi}{2}\textbf{I}_{M},
   
\end{array}\label{bar_A}
  \end{equation}
  \begin{equation}
\begin{array}{ll}
   \displaystyle \bar{\textbf{c}}_k=\sum\nolimits_{n=1}^{N_u}\varpi_{kn}\sqrt{(1+\gamma_{kn})}\textbf{a}_{kn}+\frac{\xi}{2}(\textbf{q}_k+\textbf{u}_k),
   
\end{array}\label{bar_c}
  \end{equation}
where $\textbf{q}_k=[q_{1k},...,q_{Mk}]^T\in\mathbb{R}^{M\times 1},~\textbf{u}_k=[u_{1k},...,u_{Mk}]^T\in\mathbb{R}^{M\times 1},~\forall k$. Subsequently, we can obtain the optimal update of $\boldsymbol{q}$ by solving the following optimization
problem:
 \begin{equation}
\begin{array}{ll}
   \displaystyle \boldsymbol{q}&\displaystyle\gets\text{arg}\min\limits_{\boldsymbol{q}}\mathbb{I}_{\mathcal{C}}(\boldsymbol{q})+\frac{\xi}{2}\sum\nolimits_{m=1}^K\sum\nolimits_{k=1}^K(q_{mk}-\zeta_{mk}+u_{mk})^2.
   
\end{array}\label{q_update}
  \end{equation} 
   \begin{algorithm}[t!]
   \caption{ADMM-based FP algorithm}
\begin{algorithmic}[1]
\renewcommand{\algorithmicrequire}{\textbf{Inputs:}}
\Require
$\epsilon_\text{FP}>0$ (FP tolerance), $\epsilon_\text{ADMM}>0$ (ADMM tolerance), $\xi>0$, iteration index $i\gets 0$; 
\State Initialize $\zeta_{mk}^{0},~\forall m,~k$.
\Repeat
\State Update $\gamma_{kn}^{(i+1)}=\frac{\displaystyle A_{kn}(\boldsymbol{\zeta}^{(i)})}{\displaystyle B_{kn}(\boldsymbol{\zeta}^{(i)})},~\forall k, n$.
\State Update $\varpi_{kn}^{(i+1)}=\frac{\displaystyle \sqrt{(1+\gamma_{kn}^{(i+1)})A_{kn}(\boldsymbol{\zeta}^{(i)})}}{\displaystyle A_{kn}(\boldsymbol{\zeta}^{(i)})+B_{kn}(\boldsymbol{\zeta}^{(i)})},~\forall k,n$.
\State  Update $\zeta_{mk}^{(i+1)}$, $~\forall m,k$ with $\gamma_{kn}^{(i+1)}$ and $\varpi_{kn}^{(i+1)}$ by the following steps:
\State Initialize $q_{mk}^0,~\forall, m,k$,~$u_{mk}^0\gets 0,~\forall, m,k$, sub-iteration index $t\gets 0$; 
\Repeat
\State Update $\boldsymbol{\zeta}_k^{(t+1)}\gets\bar{\textbf{A}}_k^{-1}\bar{\textbf{c}}_k^{(t)},~\forall k$
\State where $\bar{\textbf{c}}_k^{(t)}=\sum\nolimits_{n=1}^{N_u}\varpi_{kn}\sqrt{(1+\gamma_{kn})}\textbf{a}_{kn}+\frac{\xi}{2}(\textbf{q}_k^{(t)}+\textbf{u}_k^{(t)})$.
\State Update $\boldsymbol{q}_m^{(t+1)}\gets\text{min}\left\{1,~\sqrt{\frac{p_d}{N_{ap}N_u||\boldsymbol{\nu}_m^{(t)}||^2}}
\right\}\times \boldsymbol{\nu}_m^{(t)},~\forall m$,
\State where $\boldsymbol{\nu}_m^{(t)}=[(\zeta_{m1}^{(t+1)}-u_{m1}^{(t)}),...,(\zeta_{mK}^{(t+1)}-u_{mK}^{(t)})]^T$.
\State Update $u_{mk}^{(t+1)}\gets q_{mk}^{(t+1)}-\zeta_{mk}^{(t+1)}+u_{mk}^{(t)},~\forall k$.
\State Set $t \gets t+1$.
\Until 
$||\boldsymbol{\zeta}^{(t)}-\boldsymbol{q}^{(t)} ||/||\boldsymbol{\zeta}^{(t)}||\leq\epsilon_\text{ADMM}$ is satisfied.
\renewcommand{\algorithmicensure}{\textbf{Output:}}
\State ADMM-based Output:
$\zeta_{mk}^{(i+1)}=\zeta_{mk}^{(t)},~\forall m,~k$.
\State $i\gets i+1$
\Until 
$|f_1(\boldsymbol{\zeta},\boldsymbol{\gamma})^{(i)}-f_1(\boldsymbol{\zeta},\boldsymbol{\gamma})^{(i-1)}|^2\leq \epsilon_\text{FP}$ is satisfied.
\renewcommand{\algorithmicensure}{\textbf{Output:}}
\Ensure
The global optima: $\zeta_{mk}^{\text{opt}}=\zeta_{mk}^{(i)},~\forall m,~k$.
\end{algorithmic}
\label{Algorithm_FP}
 \end{algorithm}
Then, \eqref{q_update} can be decomposed as $M$ independent projections on $L_2$-norm balls:
 \begin{subequations}
  \begin{align}
   &\mathop {\min}\limits_{q_{mk}}\frac{\xi}{2}\sum\nolimits_{k=1}^K(q_{mk}-\zeta_{mk}+u_{mk})^2.
   \\
   & \text{subject~to} \nonumber \\
   & \sum\nolimits_{k=1}^K q_{mk}^2 \leq \frac{p_d}{N_{ap}N_u},~\forall m.
  \end{align}
  \label{GA_optimization5}
 \end{subequations}
By using the Karush-Kuhn-Tucker conditions for each subproblem \cite{9293031,7874154}, the optimal solution of \eqref{GA_optimization5} can be expressed as
\begin{equation}
\begin{array}{ll}
   \displaystyle \textbf{q}_m=\text{min}\left\{1,~\sqrt{\frac{p_d}{N_{ap}N_u||\boldsymbol{\nu}_m||^2}}
\right\}\times \boldsymbol{\nu}_m,~\forall m,
\end{array}\label{q_optimal}
  \end{equation} 
 where $\textbf{q}_m=[q_{m1},...,q_{mK}]^T\in\mathbb{R}^{K\times 1}$ and $\boldsymbol{\nu}_m=[(\zeta_{m1}-u_{m1}),...,(\zeta_{mK}-u_{mK})]^T\in\mathbb{R}^{K\times 1}$.  
 Thus, the proposed ADMM-based FP algorithm for downlink sum SE maximization can be stated in Algorithm \eqref{Algorithm_FP} at the previous page, where the ADMM algorithm to solve \eqref{GA_optimization4}, by updating $\boldsymbol{\zeta},~\boldsymbol{q},~\boldsymbol{u}$, is a substep algorithm in Algorithm \eqref{Algorithm_FP}. Then, the optimal power control coefficients can be delivered by  $\eta_{mk}^{\text{opt}}=\left(\zeta_{mk}^{\text{opt}}\right)^2/z_{mk},~\forall m,k$.

 \addtolength{\topmargin}{0.05in}
\subsection{Computational Complexity Analysis}
To measure the feasibility of the proposed algorithm, we introduce the computational complexity analysis of Algorithm 1\cite{9293031}. We can find that to take the inverse of the $M\times M$ matrix $K$ times in \eqref{zeta_update2} is the most computationally intensive operation for the ADMM algorithm in Algorithm 1
\cite{7874154,9293031}. Then, we approximate the computational complexity of the proposed ADMM-based FP algorithm as $O(M^3K+MN_{ap}K^2N_u^2+KN_uL)$ \cite{9293031,10812717}.
 
\vspace{-10pt}
\section{Numerical Results}
This section provides performance analysis and illustrates the proposed ADMM-based FP power allocation algorithm.
We utilize a two-dimensional coordinate system akin to \cite{10297571,44444}. The STAR-RIS is located at $(x^{\text{STAR-RIS}},y^{\text{STAR-RIS}})=(0.5,0.5)$, APs are randomly distributed with $x^{\text{AP}},y^{\text{AP}}\in[-0.5,0.5)$. Reflection-area users are located with $x^{\text{user}}\in\left[0,0.8\right]$
and $y^{\text{user}}\in\left[0,0.5\right)$, while transmission-area users are distributed with $x^{\text{user}}\in\left[0,0.8\right]$
and $y^{\text{user}}\in(0.5,0.8]$. All geographic settings are in kilometers\cite{10297571,44444}. We utilize the path loss model in \cite{7827017} with its settings to obtain the large-scale fading coefficients. Unless stated, the length of coherence block is $\tau_c=200$ symbols, $\tau_p=KN_u/2$, $p_{p}=20~\text{dBm},~p_{d}=23~\text{dBm}$, $\sigma^2=-96$ dBm, and $d_h=d_v=\lambda/4$ for the STAR-RIS elements. $\epsilon_\text{FP}=\epsilon_\text{ADMM}=0.01$, $\xi=0.1$ are set for the ADMM-FP algorithm. For comparison, $\eta_{mk}=\frac{p_d}{K\text{tr}(\hat{\mathbf{\Delta}}_{mk})},~\forall m,~k$ is treated as the no-power-control scenario and the fractional power control with $0\leq\alpha\leq1$ is introduced by\cite{44444}
\begin{equation}
\begin{array}{ll}
     \displaystyle \eta_{mk}=\left({\left(\sum\nolimits_{m'=1}^M\text{tr}(\mathbf{\Delta}_{m'k})\right)^{\alpha}\sum\nolimits_{k'=1}^K\frac{\text{tr}(\hat{\mathbf{\Delta}}_{mk'})}{\left(\sum\nolimits_{m'=1}^M\text{tr}(\mathbf{\Delta}_{m'k'})\right)^{\alpha}}}\right)^{-1}~\forall m,~k.
\end{array}\label{downlink_power_control}
   \end{equation}
 
   \begin{figure}[t!]
		\centering
	\includegraphics[width=0.76\columnwidth]{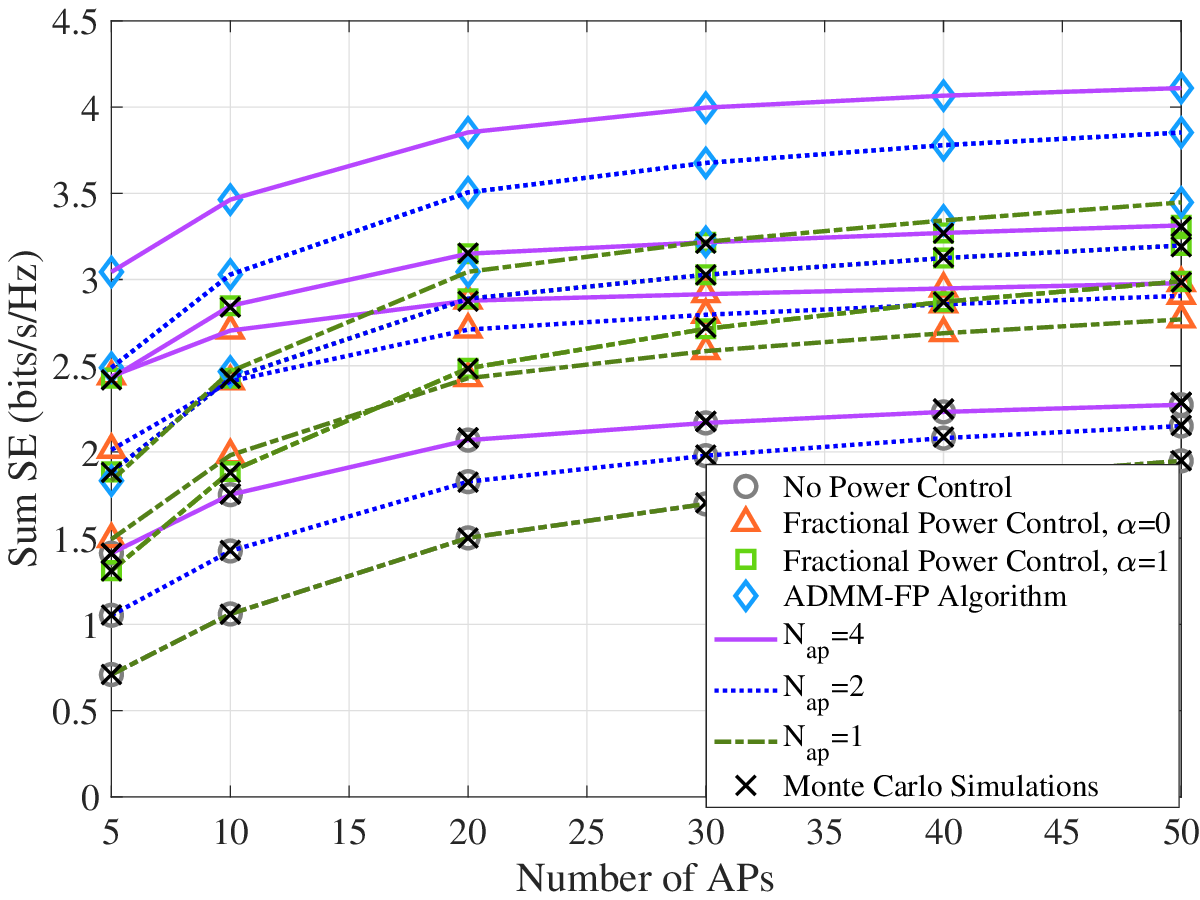} 
        \caption{Sum SE vs Number of APs with $L=16$, $K=10$, $K_t=K_r=5$, $N_u=4$.}
		\label{fig_2}
          \vspace{-7pt}
        \end{figure}
        \begin{figure}[t!]
		\centering
		\includegraphics[width=0.76\columnwidth]{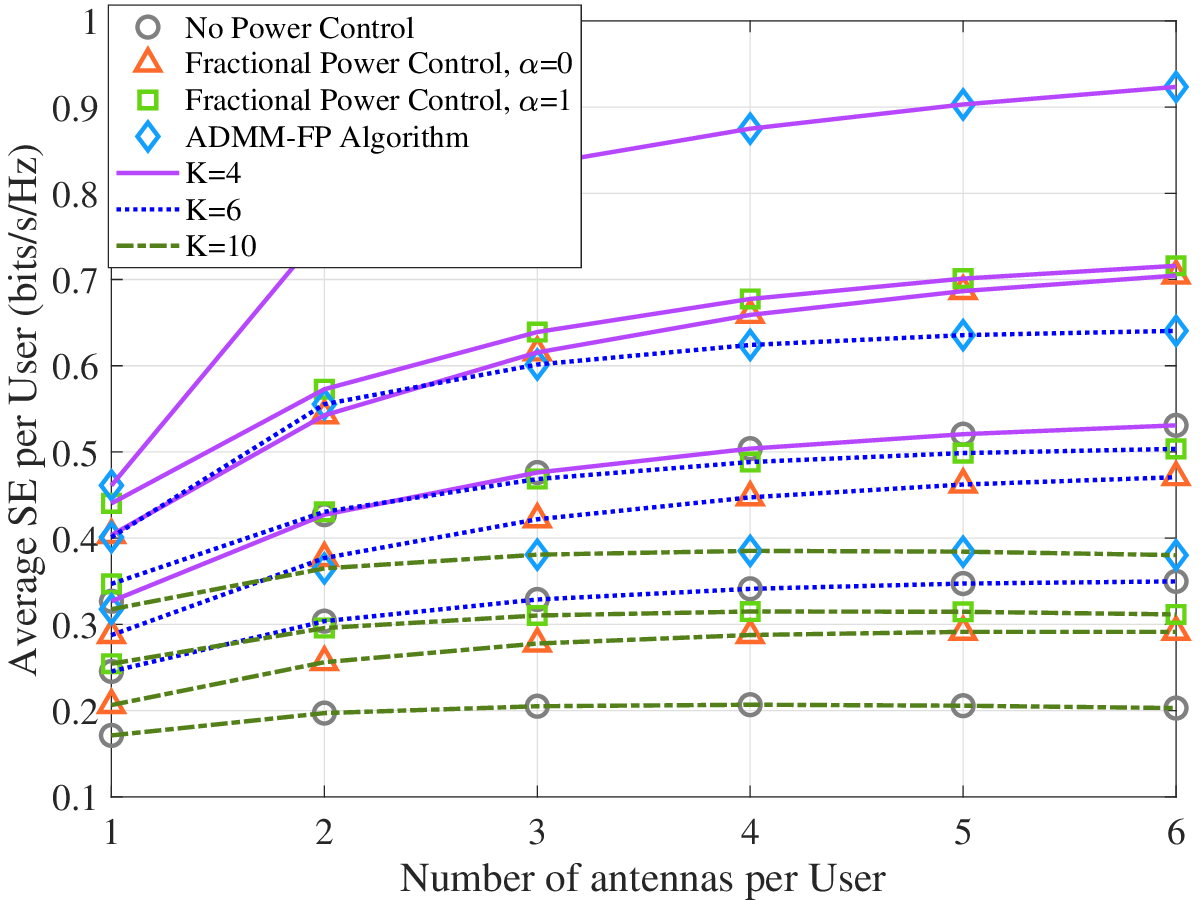} 
        \caption{Average SE vs Number of antennas per User with $M=20$, $N_{ap}=4$, $L=16$, $K_t=K_r=K/2$.}
		\label{fig_3}
          \vspace{-7pt}
        \end{figure}
        \begin{figure}[t!]
		\centering
\includegraphics[width=0.76\columnwidth]{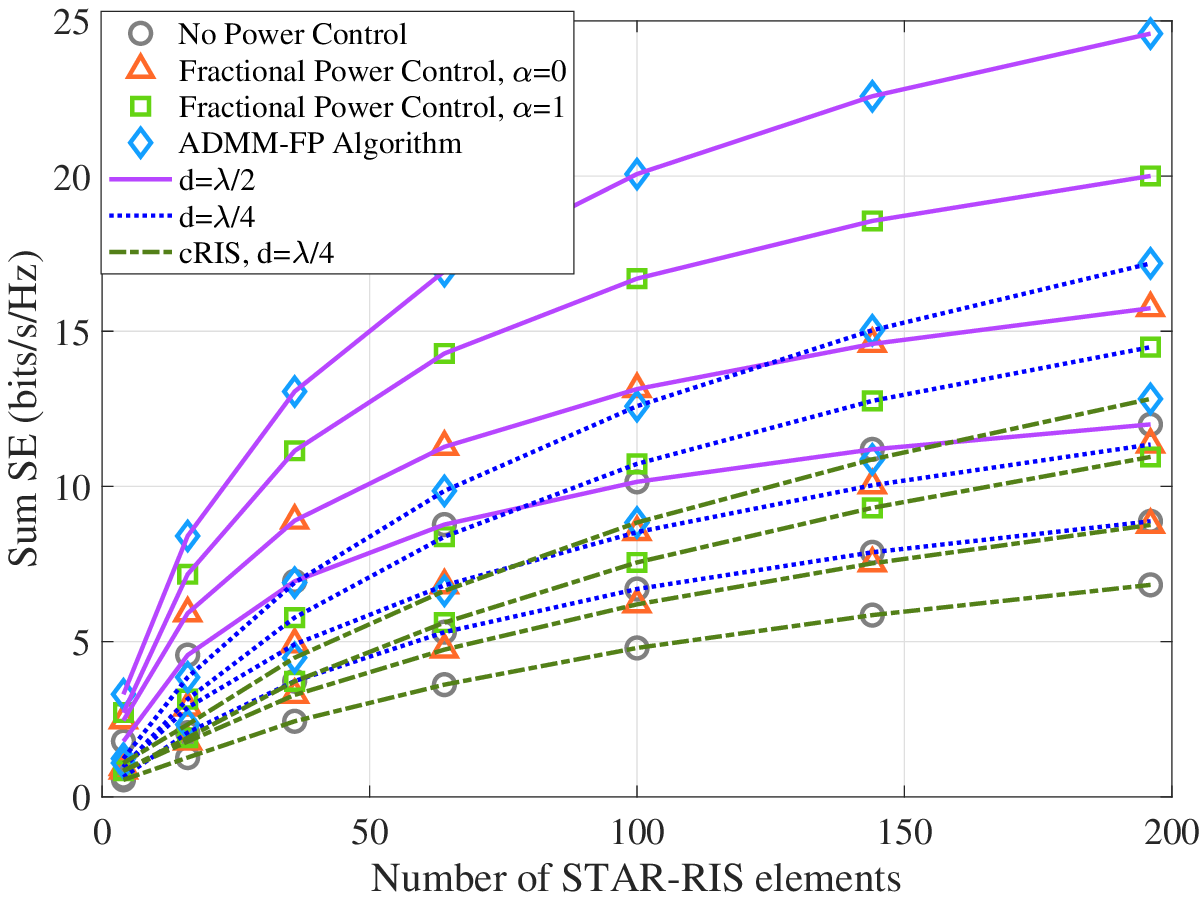} 
		\caption{Sum SE vs Number of STAR-RIS elements with $M=20$, $N_{ap}=4$, $K=10$, $K_t=K_r=5$, $N_u=4$.}
		\label{fig_4}
  \vspace{-7pt}
\end{figure}

Fig. \ref{fig_2} shows the sum SE, $\frac{\tau_c-\tau_p}{\tau_c}\sum\nolimits_{k=1}^{K}\text{SE}_{k}$, as a function of the number of APs with different numbers of antennas per AP. The closed-form analytical results delivered by \eqref{Linear_detector}-\eqref{Sigma_k_closed_form} can closely match the Monte Carlo simulations. We can find that increasing the number of APs and antennas per AP, allowing for greater spatial freedom in beamforming, can significantly improve system performance. For example, $M=50$ introduces an approximately $30\%$-likely SE increase than $M=10$ when $N_{ap}=4$ and $N_{ap}=4$ can introduce more than $10\%$-likely SE increase than $N_{ap}=2$ when $M=20$. Moreover, the proposed ADMM-based FP algorithm can introduce a respective $20\%$-likely and $80\%$-likely SE increase than the fractional power control with $\alpha=1$ and the no power control scenario when $M=20$, indicating the benefits of the proposed ADMM-based FP algorithm. Thus, properly raising the number of APs and antennas per AP, facilitated with efficient power allocation schemes, is necessary to meet the required performance. 

Fig. \ref{fig_3} depicts the average SE per user, $\frac{\tau_c-\tau_p}{K\tau_c}\sum\nolimits_{k=1}^{K}\text{SE}_{k}$, against the number of antennas per user. In particular, increasing the number of antennas per user yields up to $20\%\sim100\%$-likely average SE increase compared to single-antenna users. To be more specific, six-antenna users can introduce a $100\%$-likely SE increase with the ADMM-based FP algorithm and an over $50\%$-likely SE increase with the fractional power control than single-antenna users when $K=4$, highlighting the necessity of introducing multi-antenna users. However, this benefit decreases with a larger number of users. Meanwhile, the proposed ADMM-based FP algorithm introduces a more than $25\%$-likely average SE increase compared to the fractional power control with $\alpha=1$ when $N_u=6$. Nevertheless, the average performance diminishes with increasing users due to higher pilot contamination and inter-user interference. Thus, leveraging advanced channel estimation schemes and multi-user interference elimination schemes becomes critical to serve more users efficiently. Although the ADMM-based FP algorithm performs much better, it takes a longer running time, e.g., the ADMM-based FP algorithm takes 1.5 to 3 seconds and the fractional power control in \eqref{downlink_power_control} takes 0.5 to 1 second when $K=6$ and $K=10$, respectively. Thus, with a moderate number of users, the ADMM-based FP algorithm is recommended. However, when there is a huge quantity of users or the practical reliability requirement is not stringent, fractional power control or power allocation algorithms with much lower complexity should be introduced. 

Fig. \ref{fig_4} presents the sum SE relative to the number of STAR-RIS elements. For comparison, we introduce the scenario where a reflecting-only RIS and a conventional transmitting-only RIS are located next to each other at the STAR-RIS location. Each conventional RIS (cRIS) comprises $L/2$ elements for fairness \cite{9570143}.
The results reveal that increasing STAR-RIS elements considerably improves the sum SE. Meanwhile, a larger inter-antenna distance reduces the spatial correlation of the STAR-RIS, introducing over a $35\%$-likely SE increase when $L=196$, consistent with the results in\cite{44444}. 
Compared to the cRIS-assisted cell-free massive MIMO, the incorporation of STAR-RIS achieves a $30\%$-likely SE increase when $L=196$, highlighting the importance of STAR-RIS deployment in cell-free massive MIMO systems to boost multi-antenna system performance.

\section{Conclusion}
This work analyzed STAR-RIS-assisted cell-free massive MIMO with multi-antenna users to reveal the downlink SE with linear MMSE detectors, featuring novel closed-form SE derivations. We proposed an ADMM-based FP power control algorithm to maximize the downlink sum SE. Our numerical results indicate that six-antenna users can achieve at least a $20\%$-likely SE increase compared to single-antenna configurations, highlighting the benefits of advanced antenna technologies. Also, the ADMM-based FP algorithm outperforms fractional power control by over a $20\%$-likely SE increase. These findings emphasize the necessity of efficient power allocation algorithms and the advantages of introducing multi-antenna users and STAR-RISs in future wireless communication networks. 
\ifCLASSOPTIONcaptionsoff
  \newpage
\fi


%




\ifCLASSOPTIONcaptionsoff
  \newpage
\fi



\bibliographystyle{IEEEtran}
\bibliography{IEEEabrv,ref}
\end{document}